\documentclass[letterpaper,twocolumn,10pt]{article}
\usepackage{usenix}

\usepackage{tikz}
\usepackage{amsmath}
\usepackage{booktabs}
\usepackage{adjustbox}
\usepackage{pdflscape}
\usepackage{longtable}
\usepackage{caption}
\usepackage{listings}
\usepackage{xcolor}
\usepackage{url}

\begin{document}

\date{}

\title{\Large \bf SCION Path Performance Toolkit and Benchmark for Advancing Machine Learning in Next-Generation Networks: ScionPathML}

\author{
{\rm Damien Rossi}\\
CESI Engineering school,\\ Toulouse, France 
\and
{\rm Lars Herschbach}\\
Goethe University Franckfurt,\\ Franckfurt, Germany
\and
{\rm Sina Keshvadi}\\
Thompson Rivers University,\\ Kamloops, Canada
} 

\maketitle

\begin{abstract}
Machine learning (ML) holds great promise for optimizing networks. However, applying it to new architectures such as SCION is challenging due to a lack of relevant, accessible network performance data. This paper introduces the SCION Path Performance Toolkit (ScionPathML), a Python library designed to simplify the collection of SCION path characteristics (e.g., RTT, loss, availability) in ML-friendly formats. ScionPathML abstracts the complexities of SCION's underlying tools, enabling researchers to easily generate time-series datasets from environments like SCIONLab. Furthermore, we propose a standardized benchmarking framework, including initial datasets and tasks (e.g., RTT prediction), to foster reproducible research. This aims to track progress in the application in ML to SCION. Our contributions aim to lower the entry barrier for ML researchers and accelerate innovation in multipath transport optimization over SCION.\\ 
\end{abstract}


\section{Introduction}
The emergence of path-aware network architectures, such as SCION (Scalability, Control, and Isolation On Next-generation networks), signifies a paradigm shift in controlling, monitoring, and optimizing networks. SCION provides end-hosts with unprecedented capabilities in path control, multipathing, and offers strong security guarantees. These attributes naturally lend themselves to data-driven optimization using machine learning (ML). However, the SCION ecosystem currently lacks the tools and datasets necessary to facilitate such research.

In this paper, we present ScionPathML, an open-source Python library that attempts to close this gap. ScionPathML makes it easy to find SCION paths and obtain performance measurements over time, including end-to-end latency, packet loss, bandwidth, jitter, and per-hop RTTs. Beyond data collection, the toolkit also makes it easy to convert measurements into ML-friendly formats suitable for predictive modeling, anomaly detection, and performance diagnosis.

We intend to stimulate reproducible machine learning research into path-aware networking via the provision of (I) a robust and reusable software kit, (II) the first comprehensive public SCION path performance dataset, and (III) benchmark tasks and baseline models for traffic engineering, security, and diagnostic use cases.

\section{Background and Motivation}
Next-generation Internet architectures like SCION promise major improvements in routing security, availability, and path control. Traffic flows in SCION can traverse multiple network paths, with end-hosts inspecting all paths before making a choice that has cryptographic guarantees. This makes a number of use cases possible, including high availability, user-customization of routing based on preferences, and secure communication in the presence of a failure.

A major obstacle to this vision has been the difficulty of collecting SCION performance data at scale. Running multiple measurement tools across diverse AS pairs requires careful coordination, consistent scheduling, and robust error handling —- tasks that are error-prone and time-consuming for researchers unfamiliar with SCION’s operational details.

ScionPathML addresses these challenges by integrating core SCION measurement utilities into a unified, automated pipeline. The framework standardizes data collection, ensures consistent metadata, and outputs results in formats suitable for machine learning workflows. In this study, ScionPathML is used not as the subject of evaluation, but as the enabling infrastructure that provides the dataset for our analysis of SCION network predictability.

\section{Related Work}
Machine learning has been increasingly applied to networking problems, including traffic prediction, anomaly detection, and path optimization. Toolkits like nPrintML [4] focus on packet-level feature extraction for ML models, while projects such as RIPE Atlas [5] provide large-scale Internet measurement capabilities but lack direct SCION integration. Similarly, iPerf and its derivatives [6] offer throughput measurements, but require manual orchestration and do not natively support multipath or SCION path-aware measurements.

Research in path-aware networking has highlighted the need for richer datasets and standardized benchmarks. For example, P4-NetML [7] integrates programmable data plane measurements with ML pipelines, but is targeted at SDN/P4 environments rather than SCION.

In contrast, ScionPathML is designed specifically for the SCION architecture, integrating multiple SCION CLI tools into a unified Python library for automated, reproducible, and ML-ready dataset generation. Unlike existing toolkits, ScionPathML not only collects performance metrics (RTT, loss, jitter, bandwidth) but also offers a standardized benchmarking framework with baseline tasks, enabling direct comparison across research efforts.

\section{Toolkit Design}
ScionPathML was developed with a single goal in mind: to facilitate access to machine learning research in next-generation path-sensitive networks, such as SCION. As such, the toolkit has been organized to meet three fundamental objectives:
\\

\textbf{Develop a Toolkit:} Create and publish a reusable open-source Python library that automatically collects performance metrics from paths between AS pairs over time. Specifically, ScionPathML wraps the core SCION CLI tools of \texttt{scion showpaths},  \texttt{ scion ping}, \texttt{ scion-bwtestclient} and \texttt{ scion traceroute} to measure end-to-end latency (RTT), loss, jitter, available bandwidth, and per-hop RTTs across multiple paths in a continuous manner. Furthermore, the toolkit is designed to be simple, reusable, modular, and can be deployed on SCIONLab nodes simply and with little configuration.

\textbf{Generate a Dataset:} Using this toolkit, it is possible to collect a significant amount of data on the overall performance of the SCION network. The tool can be deployed on different ASes around the globe to collect metrics from various locations and build a robust dataset. By defining a frequency in the ScionPathML configuration, a script is executed and interacts with the other ASes and servers also added in the configuration. For each AS pair, ScionPathML can execute all supported measurement operations \texttt{showpaths, ping, bandwidth, traceroute, }allowing comprehensive information to be collected about the path and its performance profile.

This design produces a dense, time-indexed dataset of path metrics for all possible AS-AS pairs defined in the configuration. Depending on the configuration, in cases where multiple ASes are deployed with the ScionPathML tool installed on them, it is possible to cross-reference metrics at the spatial level to obtain interesting results. All measurements are stored in structured JSON files with various metadata applied (e.g., timestamp, source/destination AS, path fingerprint). Then it is possible to transform these data into CSV format so that they are suitable for use with ML.

\textbf{Benchmark:} ScionPathML has a collection of benchmark machine learning tasks, each of which is instantiated around a particular networking use case: performance prediction, failure prediction, malicious detection, path recommendation, and bottleneck localization. Each benchmark task identifies a network issue, defines a solution using the algorithm model to be used, and provides evaluation metrics to determine the quality of the results. Benchmarks are established using classic machine learning algorithm models such as Random Forest and linear regression, and typical evaluation metrics, including MAE, F1-score, AUC-ROC, and QoE satisfaction.

\section{Using ScionPathML}

\subsection{ScionPathML Workflow and Data Collection Process}

ScionPathML implements a systematic four-stage methodology for SCION network measurement and ML-ready dataset generation. The workflow progresses from the initial network configuration through automated data collection to the final preparation of the dataset for machine learning applications.

\subsubsection{Stage 1: Network Infrastructure Configuration}

The measurement process begins with a systematic configuration of the SCION measurement infrastructure. Researchers define their measurement topology by configuring autonomous systems and bandwidth testing servers.

\textbf{AS Configuration:} Researchers register each SCION autonomous system that will participate in measurements, specifying the AS ID (in standard ISD-AS format), IP address, and descriptive name. The system validates SCION addressing formats and maintains a centralized configuration registry.
If researchers need to, they can create their own ASes on the ScionLab website and integrate them into this configuration.

\textbf{Server Setup:} To measure bandwidth, researchers set up dedicated test servers. Each server is an AS but with the option enabled on it. To use bandwidth data collection, researchers also need to set up one or more servers. To do this, they must register servers in their configurations, specifying the AS ID (in standard ISD-AS format), IP address, and descriptive name.

\subsubsection{Stage 2: Measurement Pipeline Configuration}

ScionPathML coordinates multiple SCION measurement tools through a configurable pipeline system. Researchers systematically control which measurement operations are active during data collection:
\\
\textbf{Available Measurement Types:} The system coordinates seven primary measurement categories to comprehensively evaluate network paths and performance. 

\textbf{Path Discovery} utilizes the \texttt{scion showpaths} command to identify all paths available between AS pairs. To monitor stability and routing dynamics over time, \textbf{Path Comparison} employs comparer scripts that systematically analyze current path availability against previous \texttt{scion showpaths} results. 

For throughput characterization, \textbf{Bandwidth Measurements} are performed using \texttt{scion-bwtestclient}, while \textbf{Multipath Bandwidth Testing} leverages the \texttt{mp-bandwidth} script to conduct simultaneous bandwidth tests to the same destination, assessing network quality under concurrent load conditions. 

\textbf{Latency and Connectivity} metrics are gathered using \texttt{scion ping} to measure round-trip time (RTT) and reachability. Additionally, \textbf{Multipath Latency Testing} uses the \texttt{mp-prober} script to perform concurrent ping operations, evaluating consistency in network quality during simultaneous measurements.

Finally, \textbf{Path Analysis} relies on \texttt{scion traceroute} to characterize hop-by-hop latency across network paths.
\\
\textbf{Selective Measurement Control:} Researchers can enable or disable specific measurement types based on research objectives, available network resources, or experimental design requirements. This granular control allows you to customize measurement campaigns without having to manually modify scripts; this is done directly in command lines via the library.

\subsubsection{Stage 3: Automated Data Collection}

The core data collection process operates through systematic scheduling and automated execution:

\textbf{Scheduled Execution:} Researchers establish measurement intervals (e.g., 30 minutes, hourly) through cron-based scheduling. The system integrates with the host operating system's (Linux) scheduling services to ensure consistent measurement timing across extended collection periods.

\textbf{Measurement Execution Cycle:} Each scheduled execution implements a systematic process:
\begin{enumerate}
\item Systematic execution of enabled measurement commands across discovered paths 
\item Storage of structured measurement data in JSON format.
\item Generation of execution logs for monitoring
\end{enumerate}

\textbf{Data Storage Format:} All measurement results are stored in structured JSON format with a consistent metadata schema for each type of command. Each measurement record includes timestamps, source/destination AS information, path identifiers, measurement parameters, and complete results of measurements performed by the underlying SCION tools.

\subsubsection{Stage 4: Data Transformation and ML Preparation}

The final stage converts raw measurement data into analytical ready formats suitable for machine learning applications. ScionPathML provides dedicated commands to convert JSON measurement files to CSV format, allowing researchers to process entire measurement archives or specific time periods, with support for custom output locations. The resulting CSV files feature standardized column structures optimized for time series analysis and ML workflows, where each row represents a single measurement event with normalized timestamps, path identifiers, and all relevant performance metrics.

\subsection{Data Management and Organization}

\subsubsection{Systematic Data Organization}

ScionPathML implements a hierarchical data organization that supports both active measurement campaigns and historical analysis for comparison purposes, for example, available paths. The data is structured into logical categories: we have archives containing old files, history, which includes current files outside of showpaths (the N-1 version), and finally currently, which contains the latest files from the showpaths command. Researchers can move completed measurements to archives or other desired locations, search for specific files, and selectively delete data based on measurement categories or time periods. This is fully manageable via the command line with the ScionPathML tool.

\subsubsection{Monitoring and Quality Control}

The methodology includes systematic monitoring capabilities, which allow researchers to review measurement execution logs, verify data collection completeness, and identify potential issues in the measurement infrastructure. A multi-category logging system records pipeline execution status and measurement results.

\subsection{Research Application Workflow}

ScionPathML allows researchers to define workflows tailored to their specific needs. Researchers select their measurement topology, measurement types, and data collection schedules by defining their own configuration in the library. Data collection automates the execution of measurements for a data acquisition plan that systematically captures performance data over a significant period of time within the configured SCION infrastructure. Data preparation transforms raw measurement results into properly organized and consistently formatted datasets that provide useful metadata for training machine learning algorithms. This organization significantly reduces the technical burden of more traditional SCION measurement campaigns, allowing researchers to focus on training ML algorithms rather than on the difficult coordination of data creation and collection.

\section{Real-world use of the ScionPathML tool}
In order to test our tool and its ability to be useful in collecting data in the SCION environment, we conducted an experimental phase ourselves.
To do this, an environment had to be deployed in order to collect data, with the ultimate goal of performing five benchmark tasks. These tasks will validate the importance of this tool:

\subsection{Hardware Environment}
We deployed ScionPathML across four instances of Google Cloud Compute Engine, each located in a different Autonomous System (AS) region (Europe, North America, Switzerland, Taiwan). 
Each instance was configured as follows:

\begin{itemize}
    \item \textbf{Machine type:} e2-medium
    \item \textbf{vCPUs:} 2 (1 shared core)
    \item \textbf{Memory:} 4~GB
    \item \textbf{Persistent storage:} 100~GB balanced persistent disk + 10~GB balanced persistent disk
    \item \textbf{CPU Platform:} Intel Broadwell (Google Cloud E2)
    \item \textbf{Monthly estimated cost:} \(\approx\$35.46\) per instance
\end{itemize}

These nodes were interconnected through SCIONLab, enabling real-time multipath measurements between all AS pairs.

\subsection{Software Environment}
\begin{itemize}
    \item \textbf{Python version:} 3.10.12
    \item \textbf{Dependencies:} 
    \begin{itemize}
        \item \texttt{pandas$\geq$1.3.0, $<$3.0.0} --- Data manipulation and CSV handling
        \item \texttt{colorama$\geq$0.4.4} --- Cross-platform colored terminal text
        \item \texttt{tabulate$\geq$0.8.9} --- Pretty-print tabular data
    \end{itemize}
    \item \textbf{SCION Tools:} \texttt{scion, scion-bwtestclient }
    \item \textbf{Operating System:} Debian-based Linux (Google Cloud default image)
\end{itemize}

\subsection{Measurement Campaign}
The measurement campaign was conducted over a period of \textbf{four weeks}, with data collection scheduled every 30 minutes for all types of measurements:

\begin{itemize}
    \item \textbf{Path Discovery:} \texttt{showpaths}
    \item \textbf{Path Stability Tracking:} \texttt{comparer} (previous vs.\ current available paths)
    \item \textbf{Bandwidth Measurement:} \texttt{bandwidth} at three tiers --- 10~Mbps, 50~Mbps, 100~Mbps
    \item \textbf{Concurrent Bandwidth Measurement:} \texttt{mp\allowbreak-bandwidth} (two random paths tested simultaneously for each tier)
    \item \textbf{Latency Measurement:} \texttt{ping}
    \item \textbf{Concurrent Latency Measurement:} \texttt{mp\allowbreak-prober} (two random paths tested simultaneously)
    \item \textbf{Path Analysis:} \texttt{traceroute}
\end{itemize}

This setup produced a dense, time-indexed dataset across all AS-to-AS pairs. Each measurement cycle captured RTT, bandwidth, jitter, loss, per-hop RTT, and path fingerprints.

\section{Benchmark Tasks, Evaluation Framework and Baseline Results}

After using our ScionPathML tool in real-world conditions to collect data over a four-week period, we created tasks to establish a benchmark. The purpose of this benchmark is to promote the reproducibility of research and the systematic monitoring of progress. We conducted baseline experiments to assess the feasibility of predicting SCION network performance and reliability using machine learning models trained on measurements collected by ScionPathML. Here are the five tasks we defined and the results:

\subsection{Task 1: Path Performance Prediction}

\subsubsection{Definition}
\textbf{Objective:} Predict future end-to-end performance metrics (RTT, bandwidth) based on historical observations, allowing proactive path selection and network optimization.

\textbf{Problem Formulation:} Time-series forecasting using sliding windows of recent measurements (N=12 timestep) to predict performance at T+1.

\textbf{Evaluation:} Mean Absolute Error (MAE) to quantify the average prediction error in the metric's original units (ms or Mbps).

\subsubsection{Results}
This experiment aimed to forecast RTT and bandwidth values one timestep ahead based on a sliding window of 12 past measurements.

The results (Table 1) show that the RTT values are relatively stable over time, with linear regression achieving an MAE of 3.88 ms. Bandwidth, in contrast, exhibits greater variability and benefits from nonlinear models: LightGBM outperformed linear regression with a 21.47 Mbps MAE.

\begin{table}[h]
\centering
\caption{Baseline results for Task 1: Path Performance Prediction}
\label{tab:baseline-results}
\begin{tabular}{lcc}
\toprule
\textbf{Model} & \textbf{MAE (RTT)} & \textbf{MAE (Bandwidth)} \\
\midrule
Linear Regression & 3.878~ms & 24.078~Mbps \\
LightGBM          & ---     & 21.470~Mbps \\
\bottomrule
\end{tabular}
\end{table}

\begin{figure}[h]
    \centering
    \includegraphics[width=0.9\linewidth]{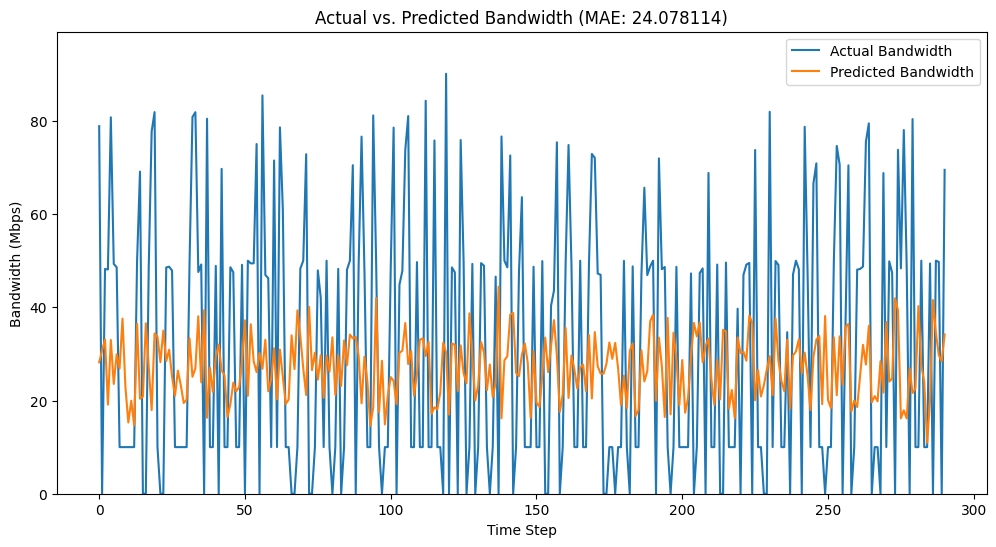} 
    \caption{Actual vs. Predicted Bandwidth}
    \label{fig:bandwidth_prediction_results}
\end{figure}

\begin{figure}[h]
    \centering
    \includegraphics[width=0.9\linewidth]{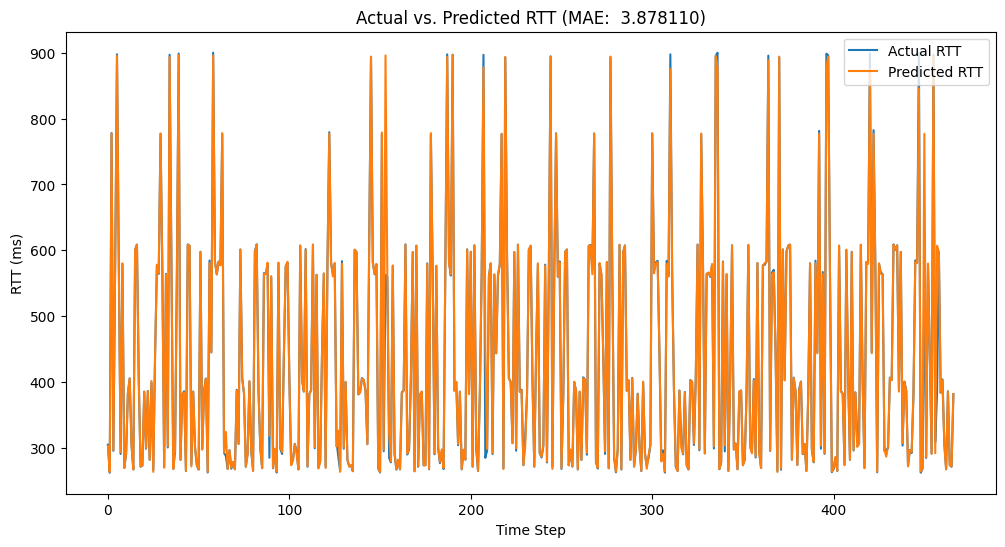} 
    \caption{Actual vs. Predicted RTT with Linear Regression Model}
    \label{fig:rtt_prediction_results}
\end{figure}

\subsubsection{Interpretation:}
This split in best-performing models highlights a potential hybrid modeling strategy, where lightweight linear predictors are used for latency metrics, and tree-based models handle more volatile bandwidth patterns.

\subsection{Task 2: Path Failure Prediction}

\subsubsection{Definition}
\textbf{Objective:} Predict impending path failures before they occur, allowing proactive traffic rerouting and improved service reliability.

\textbf{Problem Formulation:} Binary classification using time-series features extracted from recent performance measurements, incorporating trend analysis and anomaly indicators.

\textbf{Evaluation:} F1-score, precision, and recall metrics address the inherent class imbalance in failure prediction scenarios.
 
\subsubsection{Results}

The second experiment framed path failure as a binary classification problem: Given the last six RTT timesteps, jitter, loss, bandwidth, and derived deltas, predict whether a path will become unavailable at T+1.

The Random Forest model achieved an F1-score of 0.86 (Table 2). Although this indicates strong predictive power, the model occasionally missed short-lived disruptions, particularly those lasting less than a single collection interval.

\begin{table}[h]
\centering
\caption{Baseline results for Task 2: Path Failure Prediction}
\label{tab:baseline-results-2}
\begin{tabular}{lcc}
\toprule
\textbf{Model} & \textbf{F1-score} & \textbf{F1-score\%}  \\
\midrule
RandomForestClassifier & 0.860541969596 & 86 \\
\bottomrule
\end{tabular}
\end{table}

\begin{figure}[h]
    \centering
    \includegraphics[width=1\linewidth]{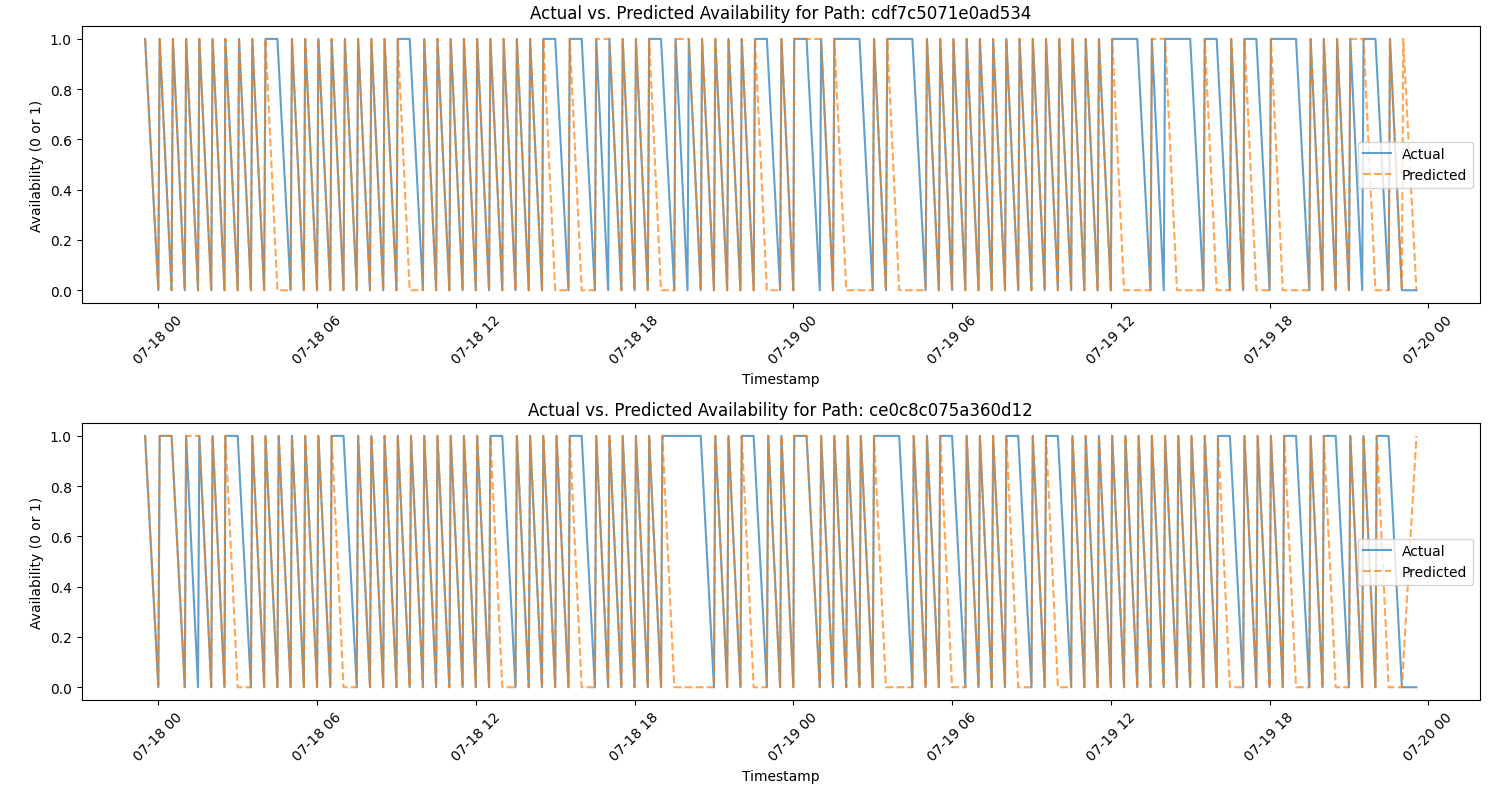} 
    \caption{Predicted Availability for Paths}
    \label{fig:path_availability_prediction_results}
\end{figure}

\subsubsection{Interpretation:}
Many SCION path failures are preceded by detectable performance declines, such as increased latency and jitter, allowing some advance warning before total loss of connectivity. However, brief interruptions and abrupt failures without any prior indicators are more difficult to anticipate, revealing limitations in current predictive monitoring approaches.  

\subsection{Task 3: Malicious Path Detection}

\subsubsection{Definition}
\textbf{Objective:} Identify paths exhibiting unusual behavior patterns that may indicate security threats, network attacks, or operational anomalies.

\textbf{Problem Formulation:} Unsupervised learning using isolation forests and autoencoders to model normal network behavior and detect statistically significant deviations.

\textbf{Evaluation:} Area Under the ROC Curve (AUC-ROC) using synthetically injected anomalies for controlled evaluation.
\subsubsection{Results}

Detecting malicious or anomalous paths is essential for maintaining the security and reliability of inter-domain networks. Unlike transient fluctuations in delay or loss, malicious behavior may manifest as persistent, structured deviations that cannot be explained by normal network dynamics. To capture such anomalies, we framed this task as an unsupervised anomaly detection problem, where the model must learn the distribution of normal path behavior and identify statistically significant deviations. In particular, we leverage isolation forests and autoencoders trained on performance time-series and per-hop traceroute features, allowing the models to capture both end-to-end and localized patterns of deviation. Ground truth anomalies were synthetically injected to enable controlled evaluation, with performance measured using the AUC-ROC metric.

\begin{table}[h]
\centering
\caption{Baseline results for Task 3: Malicious Path Detection}
\label{tab:baseline-results-3}
\begin{tabular}{lcc}
\toprule
\textbf{Model} & \textbf{AUC-ROC score} \\
\midrule
Isolation Forest & 0.774846939561274 \\
\bottomrule
\end{tabular}
\end{table}

\begin{figure}[h]
    \centering
    \includegraphics[width=1\linewidth]{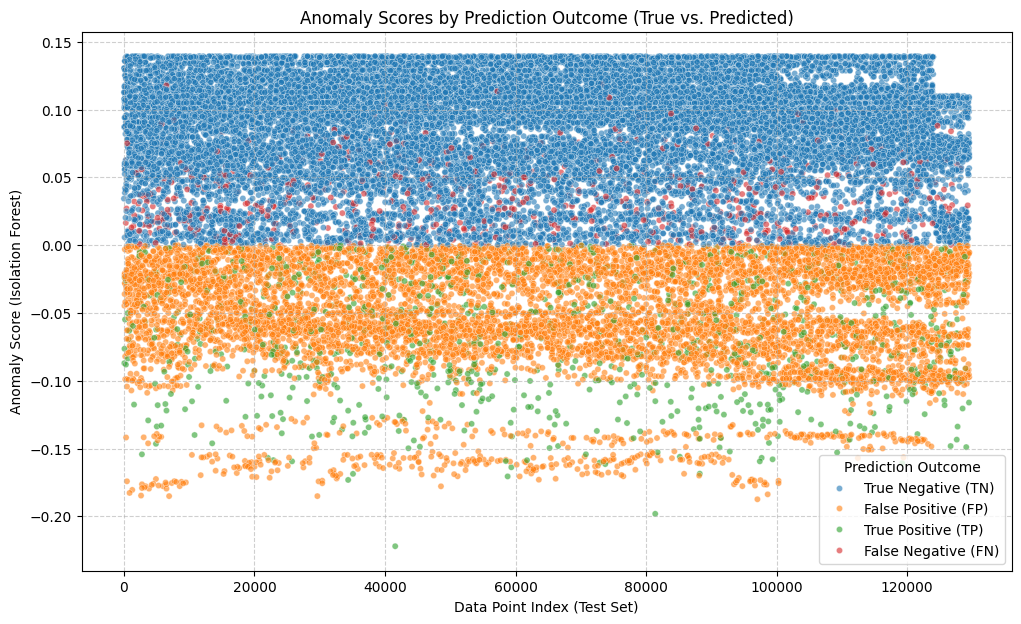} 
    \caption{Anomaly Scores by Prediction Outcome (True vs. Predicted)}
    \label{fig:anomaly_prediction_results}
\end{figure}

\subsubsection{Interpretation:}
 The baseline results indicate that the Isolation Forest achieved an AUC-ROC of  0.77, suggesting that it can distinguish between normal and anomalous paths with moderate effectiveness. This shows that unsupervised models are able to capture important deviations in path behavior even without labeled training data. However, the score also highlights limitations: some injected anomalies overlapped with the natural variability of the network, reducing separability. These findings suggest that while anomaly detection techniques provide a useful first step toward malicious path detection, more sophisticated models (e.g., deep autoencoders or ensemble methods) may be required to improve robustness against subtle or distributed attack patterns.

\subsection{Task 4: Multi-Objective Path Recommendation}

\subsubsection{Definition}
\textbf{Objective:} Given user-specified Quality of Experience (QoE) requirements, recommend optimal paths from the available set based on predicted performance.

\textbf{Problem Formulation:} Multi-criteria decision making combining predicted performance metrics with user preferences through weighted scoring functions.

\textbf{Evaluation:} QoE Satisfaction Rate measuring the percentage of recommendations that meet specified performance criteria.
\subsubsection{Results}

Modern applications place diverse and sometimes conflicting demands on the network: real-time video conferencing requires low latency and low loss, while file transfers prioritize throughput. In this task, we focus on recommending optimal paths based on user-specified Quality of Experience (QoE) profiles, which encode such requirements in terms of maximum RTT, maximum loss, and minimum bandwidth. We frame the problem as a multi-criteria decision-making task and implement a baseline heuristic approach: performance metrics are normalized, combined into a weighted score, and used to rank candidate paths. A recommendation is considered successful if the top-ranked path satisfies the corresponding QoE profile. Evaluation is based on the QoE Satisfaction Rate, i.e., the percentage of successful recommendations.

\begin{table}[h]
\centering
\caption{Defined QoE profiles}
\label{tab:QoE_Profiles}
\resizebox{\linewidth}{!}{%
\begin{tabular}{lccc}
\toprule
\textbf{Profile}    & \textbf{Max RTT (ms)} & \textbf{Max Loss (\%)} & \textbf{Min Bandwidth (Mbps)} \\
\midrule
Video conference & 150 & 2 & 1   \\
Online gaming    & 50  & 1 & 0.5 \\
File transfer    & 500 & 5 & 10  \\
Browsing         & 300 & 3 & 0.1 \\
Streaming        & 200 & 1 & 5   \\
\bottomrule
\end{tabular}%
}
\end{table}

\textbf{}

\begin{table}[h]
\centering
\caption{Baseline results for Task 4: Multi-Objective Path Recommendation}
\label{tab:baseline-results-4}
\begin{tabular}{lc}
\toprule
\textbf{QoE Profile} & \textbf{QoE Satisfaction (\%)}  \\
\midrule
Video conference & 24 \\
Online gaming    & 21 \\
File transfer    & 30 \\
Browsing         & 26 \\
Streaming        & 21 \\
\bottomrule
\end{tabular}%
\end{table}

\begin{figure}[h]
    \centering
    \includegraphics[width=1\linewidth]{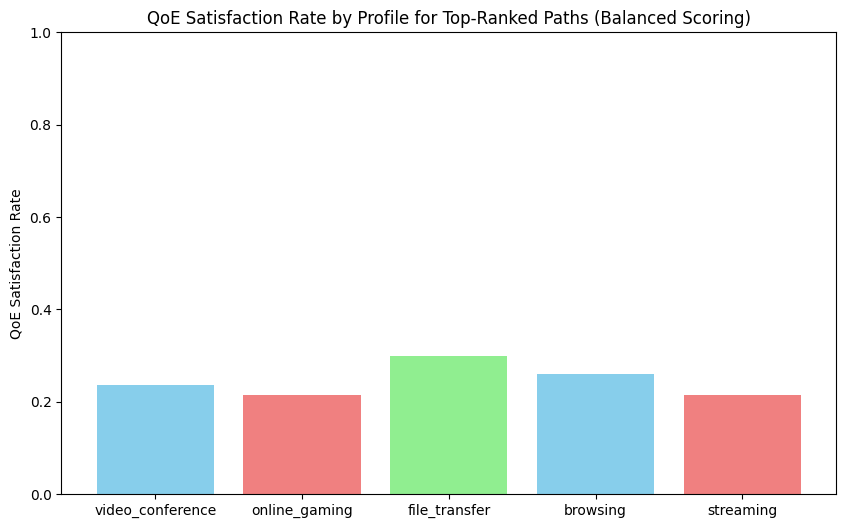} 
    \caption{Profiles QoE satisfaction results.}
    \label{fig:qoe_results}
\end{figure}

\subsubsection{Interpretation}
The heuristic baseline achieved relatively low QoE satisfaction rates across all profiles, with performance ranging between 21\% and 30\%. This reflects the difficulty of meeting strict QoE requirements in dynamic multi-path environments using a simple scoring-based strategy. In particular, stringent latency requirements (e.g., for online gaming) and high throughput demands (e.g., for file transfers) were rarely satisfied simultaneously, leading to poor recommendation reliability. These results demonstrate the need for more advanced approaches that can incorporate predictive modeling of path dynamics, adaptive weighting of QoE dimensions, and potentially reinforcement learning for online decision making.

\subsection{Task 5: Path Bottleneck Localization}
\subsubsection{Definition}
\textbf{Objective:} When path performance degrades, identify the specific network hop most likely causing the bottleneck.

\textbf{Problem Formulation:} Multi-class classification using per-hop latency vectors to predict bottleneck location.

\textbf{Evaluation:} Classification accuracy to identify synthetically introduced bottlenecks.
\subsubsection{Results}

When end-to-end path performance degrades, it is often due to congestion or faults localized to a specific hop. Accurately identifying the bottleneck hop is a prerequisite for effective remediation and troubleshooting. To address this, we framed bottleneck localization as a multi-class classification task: given a vector of per-hop latencies from traceroute at time $T$, predict the hop index where the bottleneck begins. Synthetic training data was generated by introducing artificial delays at randomly chosen hops, ensuring that the ground truth bottleneck position is known. The model is evaluated using classification accuracy.

\begin{table}[h]
\centering
\caption{Baseline results for Task 5: Path Bottleneck Localization}
\label{tab:baseline-results-5}
\begin{tabular}{lcc}
\toprule
\textbf{Model} & \textbf{Accuracy} & \textbf{Accuracy\%}  \\
\midrule
RandomForestClassifier & 0.9914 & 99 \\
\bottomrule
\end{tabular}
\end{table}

\begin{figure}[h]
    \centering
    \includegraphics[width=1\linewidth]{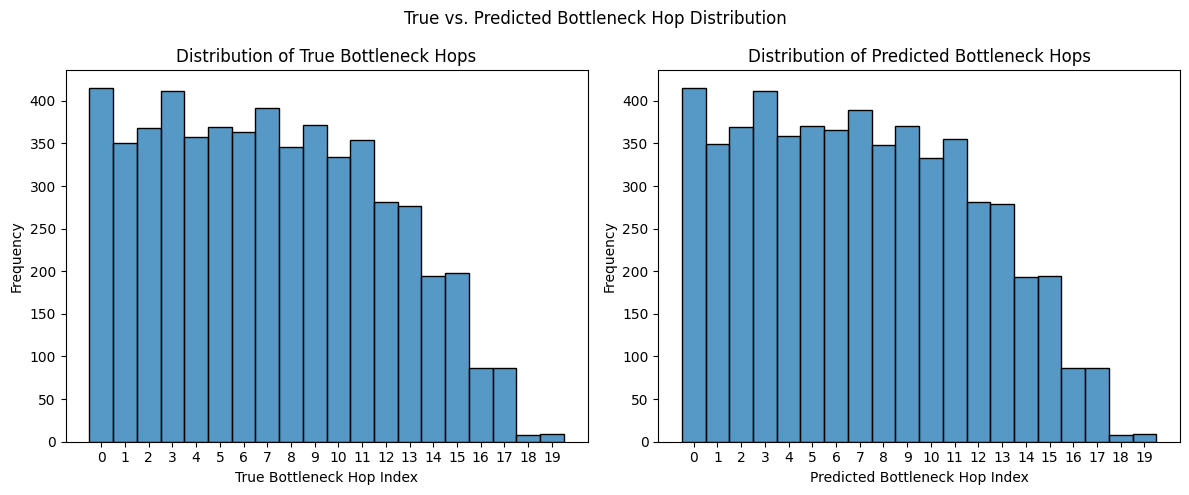} 
    \caption{True vs. Predicted Bottleneck Hop Distribution}
    \label{fig:hops_prediction_results}
\end{figure}

\subsubsection{Interpretation:}
The Random Forest classifier achieved an accuracy of 99\%, demonstrating that per-hop latency vectors provide a highly discriminative signal for localizing bottlenecks. The model was particularly effective in scenarios where bottlenecks introduced sharp, localized increases in delay, which produced clear decision boundaries. However, occasional misclassifications occurred when latency increases propagated gradually across adjacent hops, blurring the distinction between localized and distributed congestion. These results show that machine learning can reliably identify the hop most responsible for performance degradation, paving the way for automated network diagnostics in SCION and other path-aware architectures.  

\section{Discussion}

SCION datasets from baseline experiments prove to be useful for predicting both immediate performance indicators and path failures with good accuracy. Latency (RTT) displayed consistent temporal patterns that made basic models more effective at prediction, but bandwidth showed more variable patterns due to transient load conditions and multipath dynamics. The results indicate that lightweight prediction methods work well for latency forecasting, but bandwidth prediction needs more adaptable non-linear learning algorithms.

 Failure prediction analysis demonstrates that most outages occur after detectable performance drops according to an F1-score of 0.86. Our dataset's infrequent failure events require precise detection of every event because missing even one could compromise operational reliability.

A limitation of this research is its limited scope as it uses data obtained from four SCION AS nodes working under relatively consistent conditions. It is unclear if the results obtained could reasonably be assumed to exist in larger or more heterogeneous or high-churn topologies.

In general, while these initial results are indeed very promising, they only represent a first step towards predictive modeling in SCION. Moving forward, collecting data taking into account larger topologies, completing all of the benchmark tasks defined, and doing joint modeling for performance metrics and failure probabilities are important milestones for examining the viability of machine learning-based performance-aware path selection in SCION.

\section{Conclusion}
This article presented ScionPathML, a comprehensive tool that eliminates fundamental challenges to machine learning within path-aware networking. By providing easy-to-use data collection tools, standardized datasets, and clear benchmarking tasks, ScionPathML lowers access barriers to high-quality SCION performance data and advances reproducible research.

Together, our contributions represent an initial step to unlock the full potential of machine learning in path-aware networks. The toolkit provides evidence of effectiveness in lowering research barriers, improving data quality, and advancing research that was previously impossible, suggesting the critical role of community-shared infrastructure to advance research at an accelerating pace.

The standardized reference framework provides benchmarks in terms of basic performance in five functional areas for training algorithms in machine learning; therefore, this work sets clear objectives for future research while enabling and facilitating the comparison of multiple solutions. The datasets may open up new avenues of research related to network optimization and security.

As the evolution and deployment of path-aware networking architectures continue to go through adoption, tools such as ScionPathML will become increasingly essential for understanding, optimizing, and managing these complex systems. The framework is extensible to ensure that it remains relevant to future measurement and analysis techniques.

We trust that ScionPathML will stimulate the more widespread use of machine learning techniques in path-aware networking, which will ultimately result in improved efficiency, security, and reliability of next-generation Internet architectures.\\

\section*{Data Availability}

The ScionPathML toolkit is publicly available at: \url{https://github.com/Keshvadi/mpquic-on-scion-ipc/tree/ScionPathML}. \\
The Dataset is here: \url{https://github.com/Keshvadi/mpquic-on-scion-ipc/tree/main/AnalysisResults}
\\
The benchmarking tasks are here: \url{https://drive.google.com/drive/folders/1GVDPrOfT2pQe-ER_Z_YxQ1mrlRNHw89o?usp=sharing}

\end{document}